\documentclass[aip, amsmath,amssymb, reprint]{revtex4-1}
\pdfoutput=0

\usepackage{bm}
\usepackage{graphicx}
\usepackage{color}
\usepackage{dcolumn,amsmath}
\usepackage[normalem]{ulem}
\usepackage{gensymb}

\begin{document}
\newcommand{\HH}{H$_2$}
\newcommand{\pHH}{\emph{para}-H$_2$}
\newcommand{\oHH}{\emph{ortho}-H$_2$}
\newcommand{\Eref}[1]{Eq.~(\ref{#1})}
\newcommand{\Fref}[1]{Fig.~\ref{#1}}
\newcommand{\etal}{\emph{et al.}}

\title{Controlling rotational quenching rates in cold molecular collisions}

\author{J. F. E. Croft}
\affiliation{The Dodd Walls Centre for Photonic and Quantum Technologies, New Zealand}
\affiliation{Department of Physics, University of Otago, Dunedin, New Zealand}
\author{N. Balakrishnan}
\affiliation{Department of Chemistry and Biochemistry, University of Nevada,
Las Vegas, Nevada 89154, USA}
%%\email{Correspondence should be addressed to NB (naduvala@unlv.nevada.edu).}
%\author{Meng Huang}
%\author{Hua Guo}
%\affiliation{Department of Chemistry and Chemical Biology,
%University of New Mexico,
%Albuquerque, New Mexico 87131, USA}

\begin{abstract}
The relative orientation of colliding molecules plays a key role in
determining the rates of chemical processes.
Here we examine in detail a prototypical example: rotational quenching of HD
in cold collisions with \HH{}.
We show that the rotational quenching rate from $j=2 \to 0$, in the $v=1$
vibrational level, can be maximized by aligning the HD along the collision axis
and can be minimized by aligning the HD at the so called magic angle.
This follows from quite general helicity considerations and suggests that
quenching rates for other similar systems can also be controlled in
this manner.
\end{abstract}
\maketitle

\section{Introduction}
Across many areas of chemistry the dynamics of collision processes are
determined to a large extent by the relative orientation of colliding molecules
\cite{bernstein.herschbach.ea:dynamical,
levine.bernstein:molecular,
orr-ewing.zare:orientation,
orr-ewing:dynamical,
miranda.clary:quantum,
j-alexander.brouard.ea:chemistry,
aldegunde.miranda.ea:how,
jambrina.aldegunde.ea:effects}.
In the cold and ultracold regime we can study such dynamical effects with
exquisite precision
\cite{ospelkaus.ni.ea:quantum-state,
knoop.ferlaino.ea:magnetically,
perreault.mukherjee.ea:quantum,wolf.dei.ea:state-to-state,
kendrick.hazra.ea:geometric*1,croft.makrides.ea:universality,
croft.hazra.ea:symmetry,balakrishnan:perspective,bohn.rey.ea:cold}.

In a recent series of papers Perreault~\etal\ have examined how changing the
relative orientation of HD molecules in cold collisions with H$_2$ and
D$_2$ affects the angular distribution of inelastically scattered HD
\cite{perreault.mukherjee.ea:quantum,perreault.mukherjee.ea:cold}.
In the experiments HD molecules were prepared in the $v=1,j=2$ state and
the angular distribution of the scattered HD for the $v=1,j=0$ final state
was measured.
The low collision energy and light masses of the collision partners combined
with the choice of the  initial and final states limit the number of partial
waves involved in the collision process.
This was confirmed by subsequent theoretical studies which revealed that the
angular distribution of the scattered HD in collisions with~\HH\ was dominated
by a single ($l=2$) partial-wave shape resonance as around 1~K
\cite{croft.balakrishnan.ea:unraveling}.
%In the cold regime chemical processes are dominated by just a few partial waves
%which allowed for the main features of the observed angular distribution
%to be attributed to to a single ($l=2$) partial-wave shape resonance
Collisions of HD with H$_2$ are also of current astrophysical interest as the
HD molecule is believed to have played an important role in the cooling of
the primordial gas in the formation of the first stars and galaxies
\cite{galli.palla:dawn,balakrishnan.croft.ea:rotational,desrousseaux.coppola.ea:rotational}.

Here we examine how rotational quenching rates in cold molecular collisions
can be controlled by changing the relative alignment of one of the collisions
partners, taking collisions of HD with H$_2$ as a prototypical example.
We show that in order to maximize the rate of rotational quenching of HD
from $j=2 \to 0$ the HD should be aligned along the collision axis wheres to
minimize the quenching rate the HD should be aligned at the so called magic
angle.

\section{Methodology}
The H$_4$ system and its isotopologues contain just 4 electrons,
as such high quality \emph{ab initio} potential energy surfaces are
available.
In this work we have used the full-dimensional \HH-\HH\ potential of
Hinde~\cite{hinde:six-dimensional}, which is in good agreement with other
available surfaces~\cite{boothroyd.dove.ea:accurate,patkowski.cencek.ea:potential}.
Collisions of H$_2$ dimers and their isotopologues are also amenable to
quantum scattering calculations, due to the relatively shallow interaction
potential and low density of states of their energy level structure.
\cite{schaefer.meyer:theoretical,lin.guo:full-dimensional,
pogrebnya.clary:full-dimensional,
gatti.otto.ea:rotational,quemener.balakrishnan.ea:vibrational}.

Scattering calculations for collisions of HD with \HH\ were performed
in full-dimensionality using a modified version of the TwoBC code~\cite{krems}.
This methodology has been applied to many other similar systems
\cite{yang.zhang.ea:quantum,yang.zhang.ea:full-dimensional,
yang.wang.ea:full-dimensional,santos.balakrishnan.ea:vibration-vibration,
croft.balakrishnan.ea:unraveling},
and is outlined in detail elsewhere
\cite{quemener.balakrishnan.ea:vibrational,quemener.balakrishnan:quantum,
santos.balakrishnan.ea:quantum}.
Here we briefly review the methodology in order to define notation.
The scattering calculations are performed within the
time-independent close-coupling formalism which yields the scattering
$S$ matrix~\cite{arthurs.dalgarno:theory}.
For convenience, we label each asymptotic channel by the
combined molecular state $\alpha=v_1 j_1 v_2 j_2$,
where $v$ and $j$ are vibrational and rotational quantum numbers respectively,
in this work the subscript 1 refers to HD and 2 to \HH.
The integral cross section for state-to-state rovibrationally inelastic
scattering is given by,
\begin{eqnarray}
  \label{eqn:ics}
  \sigma_{\alpha \to \alpha'} = &&\frac{\pi}{(2j_1+1)(2j_2+1)k_\alpha^2} \\
  &\times& \sum_{J,j_{12},j'_{12},l,l'}(2J+1) |T^{J}_{\alpha lj_{12},\alpha'l'j'_{12}}|^{2}. \nonumber
\end{eqnarray}
where $k^2=2 \mu E/\hbar^2$ is the square of the wave vector, $E$ is the
collision energy, $\mu$ is the reduced mass, $T^J = 1-S^J$,
{\bf j$_{12}$} = {\bf j}$_1$ + {\bf j}$_2$, $l$ is the orbital angular momentum
quantum number, and $J$ the total angular momentum quantum number
{\bf J} = {\bf l} + {\bf j$_{12}$}.
To compute differential cross sections we also need the scattering
amplitude as a function of the scattering angle $\theta$,
which has previously been given by Schaefer
\etal~\cite{schaefer.meyer:theoretical} in the helicity representation:
\begin{eqnarray}
  q(\theta) = \frac{1}{2k_\alpha}&&\sum_{J}(2J+1)\sum_{j_{12},j'_{12},l,l'}i^{l-l'+1}T^J_{\alpha lj_{12},\alpha'l'j'_{12}} \nonumber \\
  &\times&d^J_{m_{12},m'_{12}}(\theta) \\
  &\times& \big\langle j'_{12}m'_{12}J-m'_{12}|l'0\big\rangle \big\langle j_{12}m_{12}J-m_{12}|l0\big\rangle \nonumber\\
  &\times& \big\langle j'_1m'_1j'_2m'_2|j'_{12}m'_{12}\big\rangle \big\langle j_1m_1j_2m_2|j_{12}m_{12}\big\rangle \nonumber
\end{eqnarray}
where $d^J_{m_{12},m'_{12}}(\theta)$ is Wigner's reduced rotation matrix.
The rovibrational helicity resolved differential cross section is then obtained
by summing over $m'_1$ and $m'_2$ and averaging over $m_1$ and $m_2$,
\begin{eqnarray}
  \label{eqn:dcs}
  \frac{d\sigma_{\alpha m_{12}\to \alpha' m'_{12}}}{d\Omega} &=& \frac{1}{(2j_1+1)(2j_2+1)} \\
  &\times&\sum_{m_1,m_2,m'_1,m'_2} |q_{\alpha,m_1,m_2,m_{12} \to \alpha',m'_1,m'_2,m'_{12}}|^{2}, \nonumber
\end{eqnarray}
where $d\Omega$ is the infinitesimal solid angle $\sin\theta d\theta d\phi$.
Helicity resolved cross sections are useful as $m_{12} \to m_{12}$ transitions
conserve $j_z$ in the body-fixed frame.
This is the assumption made in the coupled-states approximation
\cite{pack:space-fixed,mcguire.kouri:quantum,mcguire:coupled-states}
where the differential cross section is given by
\begin{eqnarray}
  \label{eqn:csa_dcs}
  \frac{d\sigma_{\alpha \to \alpha'}}{d\Omega} = \sum_{m_{12}} \frac{d\sigma_{\alpha m_{12}\to \alpha' m_{12}}}{d\Omega}.
\end{eqnarray}
The coupled-states approximation has been shown to be generally valid
for inelastic and reactive collisions away from resonances
\cite{krems.nordholm:projection-reduced,quemener.balakrishnan:cold,schaefer.meyer:theoretical}.

The cross section formulas presented so far are derived assuming that the
relative orientation of the colliding molecules is uncontrolled ---
by averaging over initial projection quantum numbers ($m_1$, $m_2$, and $m_{12}$)
and summing over final projection quantum numbers ($m'_1$, $m'_2$, and $m'_{12}$).
Here, however, we are interested in the cross section when the relative
orientation of the collision partners is controlled in the manner described
in the experiments of Perreault~\etal\
on rotational quenching of HD in cold collisions with H$_2$ and D$_2$
\cite{perreault.mukherjee.ea:quantum,perreault.mukherjee.ea:cold,
perreault.mukherjee.ea:supersonic}.
In those experiments the HD molecule was prepared in a state
$|j=2,\tilde{m}=0\rangle$
by Stark-induced adiabatic Raman passage (SARP) with the quantization axis of
$\tilde{m}$ determined by the orientation of the linear polarization of the
SARP laser \cite{zare:optical,mukherjee.perreault.ea:stark-induced}.
Choosing $\beta$ as the angle between the linear polarization of the SARP laser
and the beam velocity in the laboratory frame,
the state of a molecule prepared in a rotational state $|j,\tilde{m}\rangle$
can be expressed as
\begin{eqnarray}
  \sum_{m=-j}^{j} d^{j}_{\tilde{m},m}(\beta) |j,m\rangle,
\end{eqnarray}
in terms of projections $m$ onto the relative velocity axis.
In the work of Perreault \etal\ the HD was prepared in the $v=1,j=2$ level with
two different orientations, $\beta=0$ and $\frac{\pi}{2}$ referred to as HSARP
and VSARP respectively, where the H and V refer to the horizontal and vertical
alignment of the laser polarization respectively.
HSARP therefore corresponds to initial state $|j_1=2,m_1=0\rangle$ while
VSARP corresponds to,
\begin{eqnarray}
\sqrt{\frac{3}{8}}|j_1=2,m_1=-2\rangle&-&\frac{1}{2}|j_1=2,m_1=0\rangle \\
&+&\sqrt{\frac{3}{8}}|j_1=2,m_1=2\rangle. \nonumber
\end{eqnarray}
The helicity-resolved differential cross section for molecular collisions of
molecules prepared using the SARP method is therefore given by
\begin{eqnarray}
  \label{eqn:sarp_dcs}
  &&\frac{d\sigma_{\alpha,m_{12} \to \alpha',m'_{12}}}{d\Omega} = \frac{1}{(2j_2+1)} \\
  &&\times \sum_{m_2,m'_1,m'_2}| \sum_{m_1}d^{j_1}_{0,m_1}(\beta) \times q_{\alpha,m_1,m_2,m_{12} \to \alpha',m'_1,m'_2,m'_{12}} |^{2}. \nonumber
\end{eqnarray}
Integrating over $\phi$ and taking advantage of the cylindrical symmetry of
the problem we obtain the cross section as a function of the scattering
angle $\theta$,
\begin{eqnarray}
  \label{eqn:sarp_dcs_theta}
  &&\frac{d\sigma_{\alpha,m_{12} \to \alpha',m'_{12}}}{d\theta} = \frac{2\pi \sin \theta}{(2j_2+1)} \\
  &&\times  \sum_{m_1,m_2,m'_1,m'_2}|d^{j_1}_{0,m_1}(\beta)|^2 |q_{\alpha,m_1,m_2,m_{12} \to \alpha',m'_1,m'_2,m'_{12}} |^{2}. \nonumber
\end{eqnarray}
We can now see that the effect of orienting the collision partners is to change
the relative weighting of the helicity resolved cross sections.

\section{Results}
\begin{figure}[tb]
\centering
\includegraphics[width=0.97\columnwidth]{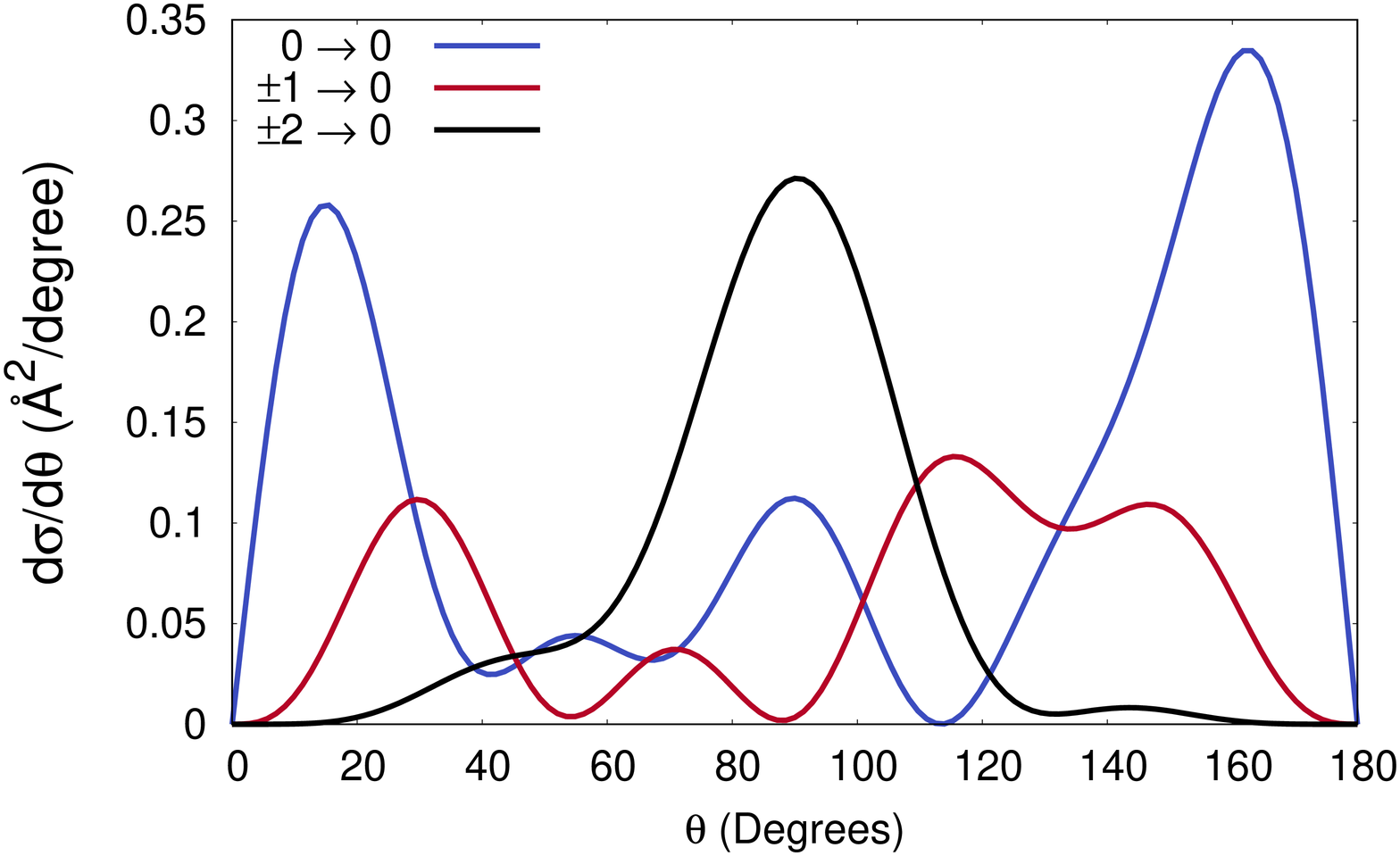}
  \caption{Helicity resolved ($m_{12} \to m'_{12}$) differential cross sections
  for rotational quenching from HD($v=1,j=2$) to HD($v=1,j=0$) in
  collisions with \pHH\ at 1~K.}
\label{fig:dcs_para}
\end{figure}
\begin{figure}[tb]
  \centering
  \includegraphics[width=0.97\columnwidth]{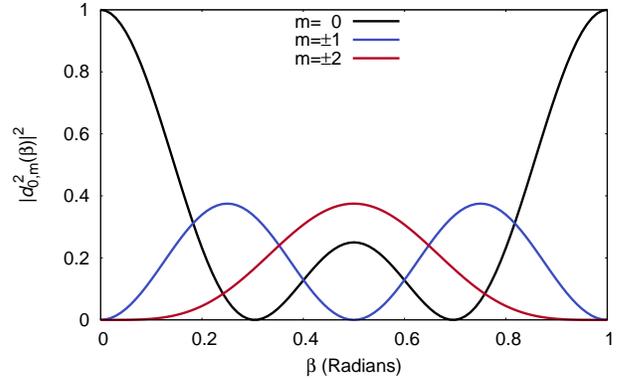}
  \caption{Wigner's reduced rotation matrix $|d^{2}_{0,m}(\beta)|^2$ as a function
  of the polar angle $\beta$ for $m=\pm2,\pm1,0$.}
  \label{fig:wigd}
\end{figure}
We first consider off resonant collisions between HD and \pHH\ due to its
relative simplicity.
As discussed helicity is approximately conserved for off resonant collisions
and the angular momentum algebra is simplified since $j_2=0$ and so
$m_{12} = m_1$.
\Fref{fig:dcs_para} shows the helicity resolved differential rate for
collisions of HD($v=1$, $j=2$) with \pHH\ at 1~K which is
off-resonant~\cite{croft.balakrishnan.ea:unraveling}.
As expected the dominant contribution is from $0 \to 0$ which conserves helicity.
In this case~\Eref{eqn:sarp_dcs_theta} is simplified as $m_2=0$ and so $m_{12} = m_1$
and $m'_1 = m'_2 = m'_{12} = 0$
\begin{eqnarray}
  \frac{d\sigma_{\alpha \to \alpha'}}{d\theta} &=& 2\pi\sin\theta \sum^2_{m_{12}=-2} |d^{2}_{0,m_{12}}(\beta)|^2 \\
  &\times& |q_{\alpha,m_1=m_{12},m_2=0,m_{12} \to \alpha',m'_1=0,m'_2=0,m'_{12}=0} |^{2}. \nonumber
%\label{eqn:dcs}
\end{eqnarray}
The effect of changing the relative alignment of the HD by $\beta$ is to
change the relative contribution from each helicity - which gives us a
handle with which to control the outcome.
For example if we were interested in maximizing the cross sections we
would want to maximize the helicity conserving contribution $0 \to 0$.
\Fref{fig:wigd} shows the square of Wigner’s reduced rotation matrix
$|d^{2}_{0,m}(\beta)|^2$ as a function of the polar angle $\beta$.
In order to maximize the cross section we choose $\beta = 0$
as the only contribution to the integral cross section that conserves
helicity ($m_{12} = m_1 = 0 \to m'_{12} = m'_1=0$) -
which corresponds to HSARP in the work of Perreault \etal\
Whereas to minimize the integral cross section we choose
$\beta = \beta_m = \arccos\frac{1}{\sqrt{3}} \approx 54.7 \degree$ the so
called magic angle, which corresponds to an initial state
\begin{eqnarray}
  \frac{1}{\sqrt{6}}|j_1=2,m_1=-2\rangle &+& \frac{1}{\sqrt{3}}|j_1=2,m_1=-1\rangle \\
  &-&\frac{1}{\sqrt{3}}|j_1=2,m_1=1\rangle \nonumber \\
  &+&\frac{1}{\sqrt{6}}|j_1=2,m_1=2\rangle \nonumber,
\end{eqnarray}
as then none of the contributions to the integral cross section conserve
helicity ($m_{12} = m_1 = \pm1, \pm2 \to m'_{12} = m'_1=0$).

\begin{figure}[tb]
\centering
\includegraphics[width=0.97\columnwidth]{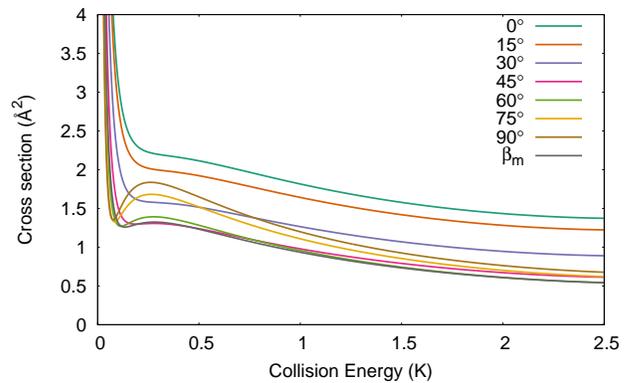}
\caption{Integral state-to-state cross sections for HD($v=1,j=2$) $\to$
  HD($v=1,j=0$) in collisions with \pHH\ for $\beta = 0,15,30,45,60,75,90$
  and the magic angle $\beta_m$}
\label{fig:ics_para}
\end{figure}

\Fref{fig:ics_para} shows the integral cross section as a function of the
collision energy for different values of $\beta$ para-H$_2$ collisions.
It is seen that as expected $\beta=0$ leads to the largest cross section
and the magic angle leads to the one of the smallest cross sections at any given
energy.
We would expect this approach to work for rotational quenching quite generally,
at least away from resonances, as the contribution from $0 \to 0$ will always
be one of the helicity conserving contributions and to work especially well
when quenching to a state where $j'_1=j'_2=0$, as we have here, where it would
be the only helicity conserving contribution.
The angle $\beta$ where $|d^{j}_{0,0}(\beta)|^2 = 0$ would however change
depending on the initial rotational state $j$ such that $P_{j}(\cos\beta) = 0$
where $P_j$ is the Legendre polynomial of degree $j$.

\begin{figure}[tb]
\centering
\includegraphics[width=0.97\columnwidth]{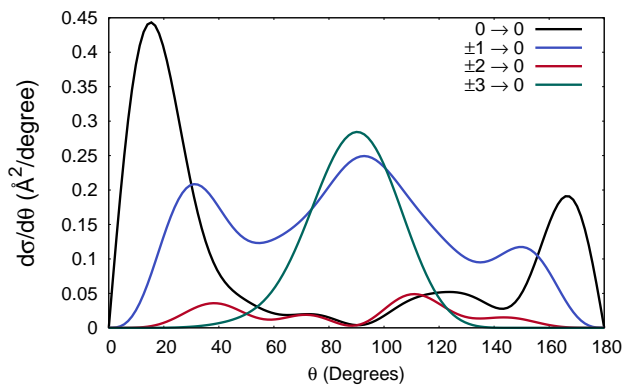}
  \caption{Dominant helicity resolved ($m_{12} \to m'_{12}$) differential
  cross sections for rotational quenching from HD($v=1,j=2$) to HD($v=1,j=0$)
  in collisions with \oHH\ at 1~K.}
\label{fig:dcs_ortho}
\end{figure}

We now move on to the more complicated case of collisions of HD with \oHH\ on
resonance.
Unlike in the off resonance case where we took advantage of the approximation
that helicity is conserved here we do no expect that to be the
case~\cite{quemener.balakrishnan:cold}.
In addition the extra angular momentum of the H$_2$ complicates the analysis.
\Fref{fig:dcs_ortho} shows the dominant helicity resolved cross sections. Unlike
in the \pHH\ case there are significant contributions to the cross section from
non-helicity conserving transitions, specifically $1 \to 0$ and $3 \to 0$.
The contribution from $2 \to 0$ on the other hand is relatively small.
Just as for the \pHH\ case $\beta$ gives a handle to vary the integral cross
section: in this case however we take advantage of the relatively small cross
section for $2 \to0$.
To maximize the cross section we again choose $\beta =0$ as $m_1=0$
can couple with $m_2 = 0,\pm1$ to make $m_{12} = m_1 + m_2 = \pm1,0$
all of which have large cross sections.
In order to minimize the cross section we want to minimize exactly those
contributions and maximize the contribution from $2\to 0$ so again we choose
the magic angle.

\begin{figure}[tb]
\centering
\includegraphics[width=0.97\columnwidth]{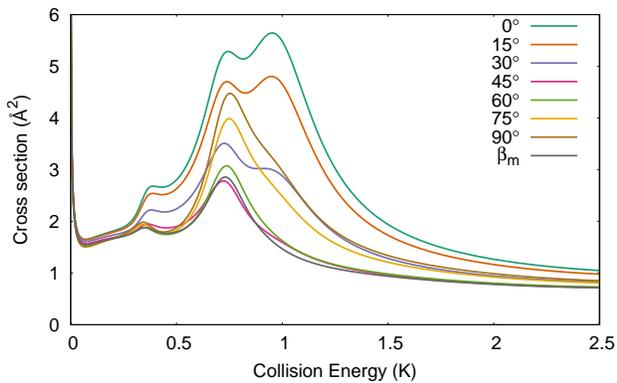}
\caption{Integral state-to-state cross sections for HD($v=1,j=2$) $\to$
  HD($v=1,j=0$) in collisions with \oHH\ for $\beta = 0,15,30,45,60,75,90$
  and the magic angle $\beta_m$}
\label{fig:ics_ortho}
\end{figure}

\Fref{fig:ics_ortho} shows the integral cross section for ortho-H$_2$ collisions
as a function of the collision energy for different values of $\beta$.
Again it is seen that, as expected, $\beta=0$ leads to the largest cross section
and the magic angle leads to the smallest cross section.
It is not clear from this case if it is in general true that the magic angle
minimizes the cross section on resonance.
We do however expect that choosing the magic angle will always reduce the
cross section compared to $\beta = 0$ as $0 \to 0$ is expected to be a
significant (even if not dominant) contribution to the total cross section
in the on resonant case.

\begin{figure}[tb]
\centering
\includegraphics[width=0.97\columnwidth]{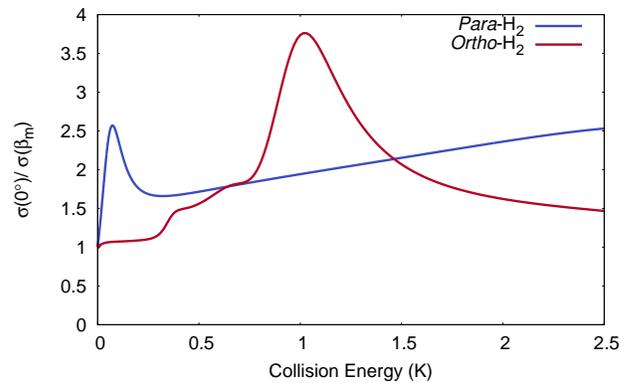}
  \caption{Ratio of the integral cross section for $\beta=0$ to
  $\beta=\beta_m$ as a function of energy.}
\label{fig:control}
\end{figure}

The degree of control over the integral cross section can be quantified
by the ratio of the integral cross section for $\beta=0$ to
$\beta=\beta_m$ $\frac{\sigma(E,\beta=0)}{\sigma(E,\beta=\beta_m)}$,
which is shown in \Fref{fig:control} as a function of collision energy.
Away from resonance it is seen that greater control over the rate of rotational
quenching can be exerted for collisions with \pHH\ than for \oHH.
This is because by choosing $\beta=0$ the only contribution to the total cross
section conserves helicity ($m_{12} = m_1 = 0 \to m'_{12} = m'_1=0$) whereas
by choosing $\beta=\beta_m$ none of the contributions conserve helicity
($m_{12} = m_1 = \pm1, \pm2 \to m'_{12} = m'_1=0$).
In collisions with \oHH\ however contributions from helicity conserving
transitions are unavoidable.
Generally speaking we therefore expect that choosing $\beta=0$ as opposed to
$\beta=\beta_m$ will have the biggest effect when quenching to a state where
$j'_1=j'_2=0$.
On the other hand the biggest effect on the integral cross section is seen
on resonance for collisions with \oHH\ where the integral cross section can be
enhanced by up to a factor of 4.
It is not clear if choosing $\beta=\beta_m$ will in general be as effective
for other resonances or whether this is a consequence of the small contribution
from $m_{12}=2 \to m'_{12} = 0$ in this case.

\section{Conclusion}
Precise stereodynamic control of cold molecular collisions has been
experimentally demonstrated by changing the relative alignment of the
collision partners using the SARP method.
In this work we have examined how rotational quenching rates in cold collisions
of HD with \HH\ can be controlled using this technique,
showing they can be enhanced by up to factor of 4:
in order to maximize the rotational quenching rate of HD
from $j=2 \to 0$ the HD should be aligned along the relative collision axis,
wheres to minimize the rate the HD should be aligned at the so called magic angle.
This follows from quite general helicity considerations and as such we expect
that rotational quenching rates for cold collisions of other similar
molecules can also be controlled in this way.
This work demonstrates the exquisite control that can be achieved in cold
molecular collisions when the incoming molecules are prepared using the SARP
method.
In future work we intend to examine rotational quenching rates for other
initial rotational states as well as other systems.

\section*{Acknowledgments}
We acknowledge support from the US National Science Foundation,
grant No.~PHY-1806334.
J.F.E.C gratefully acknowledges support from the Dodd-Walls Centre for
Photonic and Quantum Technologies.
%We thank Meng Huang, Hua Guo, Dick Zare, Nandini Mukherjee, and
%William Perreault for many stimulating discussions.
We thank Dick Zare, Nandini Mukherjee, and William Perreault for many
stimulating discussions, and Hua Guo for his careful reading of the manuscript.

\bibliographystyle{apsrev4-1}
\bibliography{../../all}

\end{document}